%% file: main.tex
\documentclass[runningheads]{llncs}
\usepackage{graphicx}
\usepackage{courier}
\usepackage[latin1]{inputenc}
\usepackage[T1]{fontenc}
\usepackage{amsmath}
\usepackage{amsfonts}
\usepackage{amssymb}
\usepackage{latexsym}
\usepackage{bussproofs}

\EnableBpAbbreviations

\usepackage{bbold} % supplies \mathbb{1}
\usepackage[dvipsnames]{xcolor} 
\definecolor{aawhite}{rgb}{0.97,0.97,0.97}
\definecolor{awhite}{rgb}{0.90,0.90,0.90}
\definecolor{lgreen}{rgb}{0.94,1.0,0.98}
\definecolor{dgreen}{rgb}{0.0,0.3,0.1}
\definecolor{sgreen}{rgb}{0.0,0.7,0.3}
\definecolor{lgreen}{rgb}{0.94,1.0,0.98}
\definecolor{bgreen}{rgb}{0.00,0.50,0.25}
\definecolor{dblue}{rgb}{0.0,0.1,0.6}
\definecolor{lblue}{rgb}{0.8,0.8,1.0}
\definecolor{mixed}{rgb}{0.0,0.3,0.3}
\definecolor{dred}{rgb}{0.6,0.2,0.0}
\definecolor{sred}{rgb}{0.7,0.2,0.0}
\definecolor{ddred}{rgb}{0.3,0.1,0.0}
\definecolor{turq}{rgb}{0.28,0.82,0.80}
\definecolor{lyellow}{rgb}{1.00,0.97,0.94}
\definecolor{mygreen}{rgb}{0,0.6,0}
\definecolor{mygray}{rgb}{0.5,0.5,0.5}
\definecolor{mymauve}{rgb}{0.58,0,0.82}
\usepackage{hyperref}
\usepackage{listings}
\input{yarel-lst.tex}
\newcommand{\verby}[1]{\lstinline[language=yarel]+#1+}
\input{java-lst.tex}
\newcommand{\verbj}[1]{\lstinline[language=java]+#1+}

\usepackage[normalem]{ulem} % supplies \sout{text}. 

\usepackage[
            backend=bibtex,
            url=false,
            doi=false,
            minbibnames=5,
            maxbibnames=5,
            defernumbers=true,
            giveninits=true
           ]{biblatex}
\bibliography{bibliography}

\begin{document}
	
\title{Introducing Yet Another REversible Language}
\author{Claudio Grandi\and
	Dariush Moshiri   \and
	Luca Roversi\orcidID{0000-0002-1871-6109}}
\authorrunning{Grandi, Claudio et al.}
\institute{Dipartimento di Informatica, C.so Svizzera 185, 10149 Torino (IT)
	\email{grandi@di.unito.it}
	\email{dariush.moshiri@edu.unito.it}
	\email{roversi@di.unito.it,luca.roversi@unito.it}
}
\maketitle      

\begin{abstract}
$ \textsf{Yarel} $ is a core reversible programming language
that implements a class of permutations, defined recursively, which are 
primitive recursive complete.
The current release of \textsf{Yarel} syntax and operational semantics, 
implemented by compiling \textsf{Yarel} to \textsf{Java}, is \texttt{0.1.0},
according to 
\href{https://semver.org/#semantic-versioning-200}{Semantic Versioning 2.0.0}. 
\textsf{Yarel} comes with 
\href{https://yarel.di.unito.it}{\textsf{Yarel-IDE}}, developed as an 
\href{https://www.eclipse.org/}{\textsf{Eclipse}} plug-in by means of 
\href{https://www.eclipse.org/Xtext/}{\textsf{XText}}.

\keywords{Reversible computation  \and Programming language \and Integrated 
development environment.}
\end{abstract}

\input{introduction}
\input{conclusions}	
\printbibliography
\end{document}

%% file: yarel-lst.tex
\lstdefinelanguage{yarel}{
    morekeywords = [1]{int, bool},
    morekeywords = [2]{ },
    morekeywords = [3]{module, import, dcl, def, dec, id, inc, neg, not,
                       inv, if, it, rec, tof},
    morekeywords = [4]{ },
    morekeywords = [5]{ },
    keywordstyle = [1]\color{blue},
    keywordstyle = [2]\color{lgreen},
    keywordstyle = [3]\color{magenta},
    keywordstyle = [4]\color{orange},
    keywordstyle = [5]\color{lblue},
    sensitive = true,
    moredelim = [s][\color{magenta}]{/}{/},
    morecomment = [l]{//},
    morecomment = [s]{/*}{*/},
    morecomment = [s]{/**}{*/},
    commentstyle={\color{dgreen}},
    morestring = [b]",
    morestring = [b]',
    literate={DOTS}{{\ldots}}1,
    basicstyle={\small\ttfamily\bfseries}, %%%% ???????
    stringstyle={\ttfamily\small\color{orange}},
    numbers=left,
    numberstyle=\tiny\color{mygray},
    xrightmargin=0em,
    xleftmargin=3em,
    stepnumber=1,
    numbersep=1em,
    lineskip=-0.5ex,
    mathescape=true,
    showstringspaces=false,
    frame=none,
    breaklines=true,
        columns=[l]{fullflexible},
    keepspaces=true,
}

\lstnewenvironment{yarel}{\lstset{language={yarel}}}{}

%% file: java-lst.tex
\lstdefinelanguage{java}{
    morekeywords = [1]{import, abstract, class, enum, extends
                      , implements, import, instanceof, interface, native
                      , new, final,  package, private, protected, public
                      , static, void},
    morekeywords = [2]{boolean, int, do, for, if, else, throws, catch, while
                      , try, null, length, assert, case, return, super, this},
    morekeywords = [3]{ },
    morekeywords = [4]{ },
    morekeywords = [5]{ },
    keywordstyle = [1]\color{magenta},
    keywordstyle = [2]\color{blue},
    keywordstyle = [3]\color{magenta},
    keywordstyle = [4]\color{orange},
    keywordstyle = [5]\color{lblue},
    sensitive = true,
    morecomment = [l]{//},
    morecomment = [s]{/*}{*/},
    morecomment = [s]{/**}{*/},
    commentstyle={\color{dgreen}},
    morestring = [b]",
    morestring = [b]',
    basicstyle={\small\ttfamily\bfseries}, %%%% ???????
    stringstyle={\ttfamily\small\color{orange}},
    numbers=left,
    numberstyle=\tiny\color{mygray},
    xrightmargin=0em,
    xleftmargin=3em,
    stepnumber=1,
    numbersep=1em,
    lineskip=-0.5ex,
    mathescape=true,
    showstringspaces=false,
    frame=none,
    breaklines=true,
    columns=[l]{fullflexible},
    keepspaces=true,
}

\lstnewenvironment{java}{\lstset{language={java},}}{}

%% file: introduction.tex
\section{Introduction}
\label{section:Introduction}
Common programming practice often depends on the solution of 
problems which are specific examples of reversible computations.
Various models that catch the meaning of reversible computation 
exist.  
One of them is the class of Reversible Primitive 
Permutations (\textsf{RPP}), introduced in 
\cite{PaoliniPiccoloRoversi-ENTCS2016,Paolini2018NGC}, 
and
simplified in \cite{PaoliniPiccoloRoversi:RPP}. Those works
speculate about how to extend the formal pattern under the design
of Primitive Recursive Functions (\textsf{PRF})
in order to capture computational reversible behaviors. 
Every permutation in \textsf{RPP} has $ \mathbb{Z}^k $ as domain 
and co-domain, for some $ k\in\mathbb{N}$.
\textsf{RPP} contains total functions and is primitive 
recursive complete, i.e. every primitive recursive function 
$ f $ can be compiled to an equivalent $ f^\bullet $ in 
\textsf{RPP} \cite{PaoliniPiccoloRoversi:RPP}.
The translation $ (\_\,)^\bullet : \textsf{PRF} \rightarrow 
\textsf{RPP}$ relies on proving that \textsf{RPP} represents 
Cantor Pairing, a pair of isomorphisms between $ \mathbb{Z}^2 $ 
and $ \mathbb{Z}^2$ that stack elements of $ \mathbb{N} $. So, 
\textsf{RPP} is at least as expressive as \textsf{PRF}.

\begin{figure}
\begin{center}
\begin{minipage}{.95\textwidth}
\begin{yarel}
module Fibonacci                                 {
 dcl coreFib : int, int
 def coreFib := /*$a$,$b$*/        it[inc]; 
                /*$a+b$,$b$*/     /2 1/; 
                /*$b$,$a+b$*/     it[inc]; 
                /*$2b+a$,$a+b$*/ inv[/2 1/]
                /*$a+b$ ,$2b+a$*/

 dcl fib : int, int, int
 def fib := /*$n \geq 0$,$a$,$b$*/ 		  /2 3 1/; 
            /*$a$,$b$,$n$*/ 		     it[coreFib]; 
            /*$a+fib(2n)$,$b+fib(2n+1)$,$n$*/ inv[/2 3 1/]
            /*$n$,$a+fib(2n)$,$b+fib(2n+1)$*/        }      
\end{yarel}
\end{minipage}
\caption{The function \verby{fib} as defined in \href{https://yarel.di.unito.it}{\textsf{Yarel-IDE}}.}
\label{figure: Yarel-IDE snapshot with Fibonacci}
\end{center}
\end{figure}
    
\paragraph{Contributions.}
\textsf{Yarel} stands for \emph{Yet Another REversible 
    Language}.
Its current release is \texttt{0.1.0}, i.e. a preliminary one,
according to 
\href{https://semver.org/#semantic-versioning-200}
{Semantic Versioning 2.0.0}.

Since \textsf{Yarel} implements \textsf{RPP}, it 
inherits its properties. Mainly, it is Primitive Recursive 
Complete. At the time of this writing we are implementing the above 
Cantor Pairing in it. Moreover, the functions we can define in \textsf{Yarel} manage their arguments linearly by construction because \textsf{RPP} builds on the monoidal structure of the algebraic theory for boolean circuits in \cite{Lafont2003257}.

\textsf{Yarel} can represent every program of \textsf{SRL} 
\cite{matos03tcs}, a reversible programming language derived 
from \emph{loop languages} 
\cite{MeyerRitchie67a}. So, \textsf{Yarel} 
inherits also the properties of \textsf{SRL}: 
(i) every \textsf{Yarel} program free of nested iterations
is equivalent to a linear
transformation $ f(x) = M x + c $, with $ M $ a matrix 
having determinant equal to 1 and $ c $ a constant 
\cite{matos03tcs};
(ii) the fixpoint problem, i.e.
``Given any function $ f\in\textsf{Yarel} $, does a tuple $ 
\overline{x} $ of values exist such that $ f(\overline{x}) 
= \overline{x} $?'', is undecidable 
\cite{PaoliniPiccoloRoversi-ENTCS2016}.

Concerning the syntax, the topmost grammatical construct of \textsf{Yarel} are the modules, every one with a name. Directives to import other 
modules, declarations of functions, i.e. their arity and types, 
and function definitions can freely alternate in a module.
Currently, the only types are comma-separated lists of the keyword \verby{int}. 
Functions in \textsf{Yarel} are the least class that we can build from
identity \verby{id}, 
increment \verby{inc}, 
decrement \verby{dec}, 
negation \verby{neg} and finite
permutations \verby{/$i_1\ldots  i_n$/}, by means of
serial composition \verby{$ f $;$ g $}, 
parallel composition\verby{$ f $|$ g $},
iteration \verby{it[$ f $]}, 
selection \verby{if[$f$,$g$,$h$]} or 
inverse \verby{inv[$ f $]}, given some function $ f, g$ and $ h $ in \textsf{Yarel}.
Figure~\ref{figure: Yarel-IDE snapshot with Fibonacci} is an example of module. Both \verby{fibCore} and \verby{fib} are translations from \textsf{SRL} \cite[p.~26]{matosWorkInProgress2014}.
Like every function in \textsf{Yarel}, \verby{fib} is arity preserving. 
The comment \verby{/*$ n\geq 0$,$ a $,$ b $*/} identifies its three 
arguments. If $ a = 0, b= 1 $, then \verby{fib}
gives triples $ (n,fib(2n), fib(2n+1)) $ by  
reorganizing its inputs in order to iterate \verby{coreFib}
$ n $ times. We exploit comments \verby{/*DOTS*/} to show the flow of the values in the functions of \textsf{Yarel}, which is \emph{point-free}, i.e. a language of combinators with no explicit reference to variable names. For example, \mbox{\verby{/*$a$,$b$*/}}\verby{it[inc];}\mbox{\verby{/*$a\,\textrm{+}\, b$,$b$*/}} in Figure~\ref{figure: Yarel-IDE snapshot with Fibonacci} says that the iteration \verby{it[inc]} of \verby{inc} maps the pair $(a,b)$ to $(a+b,b)$.

%%%%%%%%%%%%%%%%%%%%%%%%
\input{yarel-SRSOS-forward-figure}
%%%%%%%%%%%%%%%%%%%%%%%%
The operational semantics of \textsf{Yarel} is in Figure~\ref{figure:Single-relation big-step operational semantics of Yarel: first part}. Every $ \Delta $ is a list of values in $ \mathbb{Z} $, with as many elements as the arity of the function it is argument or conclusion of.  For example, the rule \textsc{itg} unfolds \verby{it[$f$]} 
to a sequential composition of
\verby{($f$|dec);DOTS;($f$|dec)}  and
\verby{($ g $|inc);DOTS;($ g $|inc)} where:
(i) the number of parallel compositions that each of them contains 
is $ v $ and
(ii) $ g $ is \verby{id|DOTS|id} with as many \verby{id} as the length of $\Delta $, i.e. the arity of $f$. 
Moreover, \textsc{itz} reduces an iteration to a suitable number of \verby{id}. Finally, $\mbox{\verby{$ \Gamma($fname$) $}}$ yields the body of the function with name \verby{fname} in \textsc{fcall} and \textsc{i-fcall}. .

\href{https://yarel.di.unito.it}{\textsf{Yarel-IDE}} is an 
integrated development environment, distributed as an \href{https://www.eclipse.org/}{\textsf{Eclipse}} plug-in,
that we generate by means of
\href{https://www.eclipse.org/Xtext/}{\textsf{XText}}, a further
\href{https://www.eclipse.org/}{\textsf{Eclipse}} plug-in 
for developing domain specific languages.
%%%%%%%%%%%%%%%%%%%%%%%%%%%%%%%%%%%%%%%%%%%%%%%%%%%%%%%%%%
\input{example-sequential-composition-compilation-figure}
%%%%%%%%%%%%%%%%%%%%%%%%%%%%%%%%%%%%%%%%%%%%%%%%%%%%%%%%%%
\href{https://www.eclipse.org/Xtext/}{\textsf{XText}} naturally leads to compile a domain specific language implemented with it into \textsf{Java} classes. \textsf{Yarel} is not an exception and the above operational semantics drives the compilation. As an example, Figure~\ref{figure:The compilation of scExample in Java}
is the object code of compiling a function \verby{seqComp}, defined as \verby{inc;dec}. The class has name \verbj{seqComp} and implements a suitable interface \verbj{RPP}.
Two private fields \verbj{l} and \verbj{r} contain the compilation of the left and of the right-hand side of the sequential composition. I.e., 
\verbj{l} is an anonymous class that contains an instance of the compilation \verbj{inc()} of \verby{inc} whose arity is in \verbj{a} and whose behaviour is in \verbj{int[] b(int[] x)}. Analogous comments hold on \verbj{r}. Given \verbj{l} and \verbj{r}, we let the arity of the sequential composition coincide with \verbj{l.getA()} while the behavior is the sequential composition of \verbj{l.b} and \verbj{r.b} in lines 15 and 16. Every function of \textsf{Yarel} is compiled under analogous patterns.
    
Remarkably, compiling some given $ f $ in \textsf{Yarel} to \textsf{Java} let
the compilation of $ f $ and of its inverse $ f^{-1} $ available in \textsf{Java} as methods that we can freely use with the proviso of never dropping any of the arguments that assure the reversibility.
We see this as pursuing the vision of \cite{DBLP:journals/tit/Huffman59}, focused on formalizing classes of classical functions with lossless inverses.
Moreover, whatever we write in \textsf{Yarel} becomes compliant with the object oriented conceptual tools that \textsf{Java} supplies without any \textit{ad-hoc} extension of \textsf{Yarel} with object oriented features.
    	
Finally, our compilation assures that the $ 32 $-bits modular arithmetic of \textsf{Java} on its integers preserves the reversibility in case of overflow. For example, a standard definition of \verby{sum} in \textsf{Yarel} compiles to a method \verbj{sum.b} whose behavior can be inverted even after computing 5 + \verbj{Integer.MAX\_VALUE}, which results in an overflow. 
We get for free what the implementation of \textsf{Janus} in~\cite[pp.~78]{DBLP:journals/entcs/Yokoyama10} requires explicitly, i.e. a sum $ u \otimes v $ on a $ 32 $-bits binary representation defined as
$ ((u+v+2^{31}) \mod 32) - 2^{31}$.

\paragraph{Related work.}
\textsf{Yarel} is functional.
Reversible \emph{functional} languages we are aware of are \textsf{RFUN}~\cite{Thomsen:2015:IPR:2897336.2897345},
\textsf{CoreFun}~\cite{DBLP:conf/rc/JacobsenKT18},
\textsf{Inv}~\cite{10.1007/978-3-540-27764-4_16} and
\textsf{Theseus}~\cite{DBLP:journals/corr/abs-1811-03678}.

The introduction of linear variables in the functional language \textsf{FUN} which leads to \textsf{RFUN} is similar to the introduction of the linear management of variables leading from Primitive Recursive Functions (\textsf{PRF}) to Reversible Primitive Permutations (\textsf{RPP}), i.e. to \textsf{Yarel}. 
A multiple output arity endows \textsf{Yarel} with an iteration \verby{it[$ f $]} and a selection \verby{if[$ f $,$ g $,$ h $]} whose inverses do not need any reference to an analogous of \textsf{RFUN}'s \emph{first-match policy}.

\textsf{Yarel} syntactically extends \textsf{SRL} \cite{matos03tcs} by means of the selection \verby{if[$ f $,$ g $,$ h $]} and of the explicit use of \verby{inv[$ f $]} that inverts the interpretation of $ f $. The operational semantics of the iteration \verby{it[$ f $]} in \textsf{Yarel} slightly differs from the corresponding construct \verbj{for x($ f $)} of \textsf{SRL}.
Every other aspects of the two languages perfectly overlap, despite \textsf{SRL} derives from loop languages~\cite{MeyerRitchie67a}. The equivalence between \textsf{Yarel} and \textsf{SRL} is open.

\textsf{Yarel} is point-less like \textsf{Inv} \cite{10.1007/978-3-540-27764-4_16} introduced to ease the management of reversible aspects in the area of document constructions. 
The basic compositional operators of \textsf{Yarel} and \textsf{Inv} overlap but differ on the basic combinators. Both languages allow the duplication of arguments, but on radical different philosophical basis. Investigating how and if the two ideas of duplication relate each other will contribute to improve our insights about the reversible computation.

The focus on type isomorphisms and on combinators that preserve information leads to \textsf{Theseus}~\cite{James2014TheseusAH}, language based on conventional pattern matching which must be subject to restrictions that guarantees no information loss. The designers of \textsf{Theseus} discarded a point-less style on purpose.
The experience of point-less programming in \textsf{Yarel} suggests to exploit
comments for a sort of correctness check. They allow to describe the flow of values that, otherwise, would remain hidden in the name of variables.

A class of circuit models coming from a categorical interpretation of the Geometry of Interaction (\textsf{GoI}) \cite{10.1007/978-3-642-79361-5_4} is in~\cite{ABRAMSKY2005441}. A possible connection is that \textsf{Yarel}, like the above classes of circuits, has a natural representation in terms of string diagrams whose computation can be described by a flow of tokens.

Also \cite{dipierro_hankin_wiklicky_2006} deals with reversible combinators. It labels combinators and encodes their reduction history, something that strongly recalls how a Turing machine becomes reversible \cite{Bennett:1973:LRC:1664562.1664568} and which we avoided since our first steps.

Finally, even though not a combinator language, we cannot forget \textsf{ROOPL}~\cite{DBLP:journals/corr/Haulund17}, object oriented extension of \textsf{Janus} \cite{Lutz86,Yokoyama:2008:PRP:1366230.1366239,paolini2018lipics}. \textsf{Yarel} is not at all an object oriented programming language, but it can interact with an object oriented environment because we compile it to \textsf{Java}.

%% file: yarel-SRSOS-forward-figure.tex
%%%%%%%%%%%%%%%%%%%%%%
\begin{figure}
	\centering
	\scalebox{.85}{
		\begin{tabular}{c}
			\AxiomC{\phantom{$ \Delta \mbox{\verby{id}}\Delta' $}}
			\RightLabel{\scriptsize \textsc{id}}
			\UnaryInfC{$  [v]\, \mbox{\verby{id}}\, [v] $}
			\DisplayProof			
			\quad
			\AxiomC{$ \Delta\,\mbox{\verby{id}}\,\Delta' $}
			\RightLabel{\scriptsize \textsc{i-id}}
			\UnaryInfC{$ \Delta\,\mbox{\verby{inv[id]}}\,\Delta' $}
			\DisplayProof			
			\quad
			\AxiomC{\phantom{$ \Delta\,\mbox{\verby{neg}}\,\Delta' $}}
			\RightLabel{\scriptsize \textsc{neg}}
			\UnaryInfC{$ [v]\,\mbox{\verby{neg}}\,[-v] $}
			\DisplayProof			
			\quad
		    \AxiomC{$\Delta\,\mbox{\verby{neg}}\,\Delta' $}
			\RightLabel{\scriptsize \textsc{i-neg}}
			\UnaryInfC{$\Delta\,\mbox{\verby{inv[neg]}}\,\Delta'$}
			\DisplayProof
			\\[0.6cm]
			\AxiomC{\phantom{$\Delta \mbox{\verby{inc}}\Delta' $}}
            \RightLabel{\scriptsize \textsc{dec}}
            \UnaryInfC{$[v]\,\mbox{\verby{dec}}\,[v-1]$}
            \DisplayProof			
			\ \
			\AxiomC{$\Delta\,\mbox{\verby{inc}}\,\Delta'$}
            \RightLabel{\scriptsize \textsc{i-dec}}
            \UnaryInfC{$\Delta\,\mbox{\verby{inv[dec]}}\,\Delta'$}
            \DisplayProof			
			\quad
			\AxiomC{\phantom{$\Delta\,\mbox{\verby{dec}}\,\Delta'$}}
			\RightLabel{\scriptsize \textsc{inc}}
			\UnaryInfC{$[v]\,\mbox{\verby{inc}}\,[v+1]$}
			\DisplayProof			
			\ \
			\AxiomC{$\Delta\,\mbox{\verby{dec}}\,\Delta'$}
			\RightLabel{\scriptsize \textsc{i-inc}}
			\UnaryInfC{$\Delta\,\mbox{\verby{inv[inc]}}\,\Delta'$}
			\DisplayProof			
			\\[0.6cm]
			\AxiomC{$\{i_1 \ldots i_n\} = \{1 \ldots n\}$}
			\RightLabel{\scriptsize $ \chi $}
			\UnaryInfC{$ [v_1, \ldots, v_n]\,
                \mbox{\verby{/$ i_1 \ldots i_n$/}}\,
				[v_{i_1}, \ldots, v_{i_n}]$}
			\DisplayProof			
			\qquad
			\AxiomC{$\{i_1 \ldots i_n\} = \{1 \ldots n\}$}
			\RightLabel{\scriptsize \textsc{i-}$\!\chi$}
			\UnaryInfC{$ [v_1, \ldots, v_n]\,
                \mbox{\verby{inv[/$i_1\ldots i_n$/]}}\, 
				[v_{i_1}, \ldots, v_{i_n}]$}
			\DisplayProof			
			\\[0.6cm]
			\AxiomC{$\Delta\,f\,\Delta'$}
			\AxiomC{$\Delta' \,g\,\Delta'' $}
			\RightLabel{\scriptsize \textsc{seq}}
			\BinaryInfC{$\Delta\,\mbox{\verby{($f$;$g$)}}\,\Delta''$}
			\DisplayProof
			\qquad
			\AxiomC{$\Delta\,g\,\Delta'$}
			\AxiomC{$\Delta'\,f\,\Delta''$}
			\RightLabel{\scriptsize \textsc{i-seq}}
			\BinaryInfC{$\Delta\,\mbox{\verby{inv[$f$;$g$]}}\, 
				 \Delta'' $}
			\DisplayProof			
			\\[0.6cm]
			\AxiomC{$\Delta_f\,f\,\Delta'_f$}
			\AxiomC{$\Delta_g\,g\,\Delta'_g $}
			\RightLabel{\scriptsize \textsc{par}}
			\BinaryInfC{$(\Delta_f \cdot \Delta_g)\,\mbox{\verby{($f$|$g$)}}\, 
				(\Delta'_f \cdot \Delta'_g)$}
			\DisplayProof			
			\qquad
			\AxiomC{$\Delta_f\,\mbox{\verby{inv[$ f $]}}\,\Delta'_f$}
			\AxiomC{$\Delta_g\,\mbox{\verby{inv[$ g $]}}\,\Delta'_g$}
			\RightLabel{\scriptsize \textsc{i-par}}
			\BinaryInfC{$(\Delta_f \cdot 
			\Delta_g)\,\mbox{\verby{inv[$f$|$g$]}}\, 
				(\Delta'_f \cdot \Delta'_g)$}
			\DisplayProof			
			\\[0.6cm]
%			\AxiomC{$ v > 0 $}
			\AxiomC{$\Delta\,g\,\Delta' $}
			\LeftLabel{\tiny $ (v > 0) $}
			\RightLabel{\scriptsize \textsc{ifg}}
			\UnaryInfC{$(\Delta \cdot [v])\,
                \mbox{\verby{if[$g$,$z$,$s$]}}\,
				(\Delta' \cdot [v])$}
			\DisplayProof			
			\qquad
%			\AxiomC{$ v < 0 $}
			\AxiomC{$\Delta\,s\,\Delta'$}
			\LeftLabel{\tiny $(v < 0) $}
			\RightLabel{\scriptsize \textsc{ifs}}
			\UnaryInfC{$(\Delta \cdot [v])\,
                \mbox{\verby{if[$g$,$z$,$s$]}}\,
				(\Delta' \cdot [v])$}
			\DisplayProof			
			\\[0.6cm]
			\AxiomC{$\Delta\,z\,\Delta'$}
			\RightLabel{\scriptsize \textsc{ifz}}
			\UnaryInfC{$(\Delta \cdot [0])\,
                \mbox{\verby{if[$g$,$z$,$s$]}}\,
				(\Delta' \cdot [0])$}
			\DisplayProof	
			\qquad
			\AxiomC{$\Delta\,
                \mbox{\verby{if[inv[$ g $],inv[$ z $],inv[$ s $]]}}\, 
				\Delta' $}
			\RightLabel{\scriptsize \textsc{i-if}}
			\UnaryInfC{$\Delta\,
                \mbox{\verby{inv[if[$ g $,$ z $,$ s $]]}}\, 
				\Delta'$}
			\DisplayProof					
			\\[0.6cm]
\AxiomC{$(\Delta \cdot [v])\,
    \mbox{\verby{($ f $|dec);it[$f$];(id|DOTS|id|inc)}}\,
	(\Delta' \cdot [v])$}
\LeftLabel{\tiny $ (v > 0)$}
\RightLabel{\scriptsize \textsc{itg}}
\UnaryInfC{$(\Delta \cdot [v])\,\mbox{\verby{it[$f$]}}\,
	(\Delta'\cdot [v])$}
\DisplayProof			
\\[0.6cm]
%\AxiomC{$ v < 0 $}
\AxiomC{$(\Delta \cdot [v])\,
	\mbox{\verby{($ f $|inc);it[$f$];(id|DOTS|id|dec)}}\,
	(\Delta' \cdot [v])$}
\LeftLabel{\tiny$  (v < 0) $}
\RightLabel{\scriptsize \textsc{its}}
\UnaryInfC{$(\Delta \cdot [v])\,\mbox{\verby{it[$f$]}}\,
	(\Delta' \cdot [v])$}
\DisplayProof
\\[0.6cm]
\AxiomC{$(\Delta \cdot [0])\,
	     \mbox{\verby{id|DOTS|id|id}}\,
	     (\Delta \cdot [0])$}
\RightLabel{\scriptsize \textsc{itz}}
\UnaryInfC{$(\Delta \cdot [0])\,\mbox{\verby{it[$f$]}}\,(\Delta'\cdot[0])$}
\DisplayProof
\qquad
\AxiomC{$\Delta\,\mbox{\verby{it[inv[$ f $]]}}\,\Delta'$}
\RightLabel{\scriptsize \textsc{i-it}}
\UnaryInfC{$\Delta\,\mbox{\verby{inv[it[$ f $]]}}\,\Delta'$}
\DisplayProof			
\\[0.6cm]
\AxiomC{$\Delta\,\mbox{\verby{$ \Gamma($fname$) $}}\,\Delta'$}
\RightLabel{\scriptsize \textsc{fcall}}
\UnaryInfC{$\Delta\,\mbox{\verby{fname}}\,\Delta'$}
\DisplayProof			
\qquad
\AxiomC{$\Delta\,\mbox{\verby{inv[$ \Gamma($fname$) $]}}\,\Delta'$}
\RightLabel{\scriptsize \textsc{i-fcall}}
\UnaryInfC{$\Delta\,\mbox{\verby{inv[fname]}}\,\Delta'$}
\DisplayProof
\qquad
\AxiomC{$\Delta\,f\,\Delta'$}
\RightLabel{\scriptsize \textsc{i-inv}}
\UnaryInfC{$\Delta\,\mbox{\verby{inv[inv[$ f $]]}}\,\Delta'$}
\DisplayProof			
\\[0.25cm]
		\end{tabular}
	}
	\caption{Operational semantics of \textsf{Yarel}. The function 
		\verby{id|DOTS|id|inc}
		in \textsc{itg} contains as many occurrences of \verby{id} as the length of $ \Delta $. An analogous comment holds for \textsc{its} and \textsc{itz}.}
	\label{figure:Single-relation big-step operational semantics of Yarel: first part}
\end{figure}

%% file: example-sequential-composition-compilation-figure.tex
\begin{figure}
\begin{java}
public class seqComp implements RPP                     {
  public seqComp() { } // Constructor
  // l(eft-hand side) of the sequential composition
  private RPP l = new RPP()                             { 
    private RPP f = new inc();  // an instance of inc
    private final int a = f.getA();
    public int[] b(int[] x) { return this.f.b(x); }
    public int getA() { return this.a; }                };  
  // r(ight-hand side) of the sequential composition
  private RPP r = new RPP()                             {
    private RPP f = new dec();  // an instance of dec
    private final int a = f.getA();
    public int[] b(int[] x) { return this.f.b(x); }
    public int getA() { return this.a; }                };
  private final int a = l.getA(); // Same arity of l or r
  public int[] b(int[] x)         { // Seq. composition
    return this.r.b(this.l.b(x)); }
  public int getA() { return this.a; }                    }
\end{java}
\caption{The compilation of \verby{scExample} in $ \textsf{Java} $.}
\label{figure:The compilation of scExample in Java}
\end{figure}

%% file: conclusions.tex
\section{Conclusions}
\label{section:Conclusions}
We are extending \textsf{Yarel} with primitive and compound types.
We are also writing a first set of libraries. Being \textsf{Yarel} a core language, everything needs to be programmed from scratch, generally producing inefficient algorithms. A way out is to first program reversible algorithms
in \textsf{Yarel} to identify the interface. Then, it is possible to re-implement them directly in \textsf{Java}, preserving that interface, but dramatically improving the efficiency. 